\documentclass[preprint,aps,shownopacs,preprintnumbers,amsmath,amssymb,
nofootinbib]{revtex4}

\usepackage{graphicx}
\usepackage{epsfig}
\usepackage{color}
\begin{document}

\begin{flushright}
\end{flushright}


\newcommand{\be}{\begin{equation}}
\newcommand{\ee}{\end{equation}}
\newcommand{\bea}{\begin{eqnarray}}
\newcommand{\eea}{\end{eqnarray}}
\newcommand{\nn}{\nonumber}
\def\CP{{\it CP}~}
\def\cp{{\it CP}}
\title{\large Non-zero $\theta_{13}$ and leptonic CP phase with $A_4$ Symmetry}

\author{Sruthilaya M.,   R. Mohanta }
\affiliation{School of Physics, University of Hyderabad, Hyderabad - 500 046, India }

\begin{abstract}
We consider a model based on $A_4$ symmetry to explain the phenomenon of neutrino mixing. The spontaneous symmetry breaking of $A_4$ symmetry 
leads to a co-bimaximal mixing matrix at leading order.
We  consider the effect of  higher order corrections in neutrino sector and find that the mixing angles thus obtained, come well within the $3\sigma$ ranges of their experimental values. 
We  study the implications of this formalism on the other phenomenological observables, such as CP violating phase, Jarlskog invariant
and the effective Majorana mass $|M_{ee}|$.  We also obtain the branching ratio of the lepton flavour violating decay $\mu\rightarrow e \gamma$ in the context of this model and find that it can be less than its present experimental upper bound. 

\end{abstract}

\maketitle
\section{Introduction}
Neutrinos are the least interacting entities among the standard model particles and exist in three flavours 
(electron neutrino, muon neutrino and tau neutrino). They change their flavour as they propagate and this 
phenomenon is known as neutrino oscillation which occurs since the flavour eigenstates of neutrinos are mixture of mass eigenstates. 
The mixing is described by PMNS matrix \cite{ri1}, 
which can be parameterized in terms of three mixing angles and three CP violating phases as
\begin{small}
\begin{equation}
V_{PMNS} = U_{PMNS}.P_\nu
 = \left( \begin{array}{ccc} c^{}_{12} c^{}_{13} & s^{}_{12}
c^{}_{13} & s^{}_{13} e^{-i\delta} \\ -s^{}_{12} c^{}_{23} -
c^{}_{12} s^{}_{13} s^{}_{23} e^{i\delta} & c^{}_{12} c^{}_{23} -
s^{}_{12} s^{}_{13} s^{}_{23} e^{i\delta} & c^{}_{13} s^{}_{23} \\
s^{}_{12} s^{}_{23} - c^{}_{12} s^{}_{13} c^{}_{23} e^{i\delta} &
-c^{}_{12} s^{}_{23} - s^{}_{12} s^{}_{13} c^{}_{23} e^{i\delta} &
c^{}_{13} c^{}_{23} \end{array} \right) P^{}_\nu \;,\label{pmns}
\end{equation}
\end{small}
where $c_{ij}=\cos\theta_{ij}$ and $s_{ij}=\sin\theta_{ij}$, $\theta_{12}$, $\theta_{23}$ and $\theta_{13}$ are the three mixing angles, 
$\delta_{CP}$ is the Dirac phase and the other two Majorana phases come in $P_{\nu}$
\begin{equation}\nonumber
P_{\nu}={\rm diag}(e^{i\rho},e^{i\sigma},1)\;.
\end{equation} 
Neutrino oscillation experiments gained a lot of interest as a probe to neutrino mixing and mass spectrum since the oscillation probability depends 
on mixing angles, Dirac CP phase and the mass square differences ($\Delta m^2_{21}$ and $\Delta m^2_{23}$).
Results from earlier experiments indicated that $\theta_{13}$ is very small, can be zero and the lepton mixing is very close to TBM 
(Tri-bimaximal mixing) \cite{ri2}, 
 which predicts $\sin\theta_{13}=0$, $\sin^2\theta_{23}={1}/{2}$ and $\tan^2\theta_{12}={1}/{2}$.  
This made it possible to explain the neutrino mixing as TBM type,  with small deviation due to perturbation in the charged-lepton or neutrino sector. 
There are many models which explain TBM mixing pattern on the basis of $A_4$  symmetry \cite{ri3} with a certain set of Higgs scalars and vacuum alignments. 
Recent experimental observations of moderately large $\theta_{13}$ \cite{th13}, made neutrino mixing a little far from TBM type, but close
to co-bimaximal mixing which predicts non-zero $\theta_{13}$ ($\theta_{13}\neq0$, $\theta_{23}={\pi}/{4}$, $\delta_{CP}=\pm{\pi}/{2}$) \cite{r7}. 
Supersymmetric models based on $A_4$ family symmetry, combined with the generalized CP symmetry \cite{A4susy}, can also predict 
trimaximal (TM) lepton mixing, (in which either only the first column or only the second column of the lepton mixing matrix is assumed to take the TBM
form), 
  together with either zero CP violation or $\delta_{CP} = \pm \pi/2$. Also models 
based on $S_4$ family symmetry and generalized CP symmetry \cite{S4sy}  predict trimaximal lepton mixing
and the  Dirac CP is predicted to be either conserved or maximally broken.
In Ref. \cite{valle}, a minimal extension of the simplest $A_4$ model has been considered, which not only can induce non-zero $\theta_{13}$
value, consistent with the recent observations, but also   can correlate the CP violation in neutrino oscillation with the 
octant of the atmospheric mixing angle $\theta_{23}$. 
In this paper, we would like to consider a model based 
on $A_4$ symmetry which gives co-bimaximal mixing in neutrino sector at leading order. To accommodate deviations in mixing angles to make them compatible with the experimental results, we include a perturbation in neutrino sector due to higher order corrections, which can be represented as five-dimensional  operators. 
 The best-fit values and $3\sigma$ ranges of neutrino oscillation parameters taken from Ref. \cite{ri4}  are given in Table \ref{t12}.

The paper is  organized as follows. The  details of our model is presented in section II. In sections III and IV, we discuss the vacuum alignment and lepton flavour violating muon decay $\mu\rightarrow e \gamma$ in the context of the model. In section V, we describe the  higher order corrections in neutrino sector and we conclude our discussion  in section VI.

\begin{table}[htb]
\begin{center}
\vspace*{0.1 true in}
\begin{tabular}{|c|c|c|}
\hline
 ~~Mixing Parameters~~ & ~~Best Fit values~~ & $~~ 3 \sigma $ Range~~  \\
\hline
$\sin^2 \theta_{12} $ &~ $0.323$ ~& ~$ 0.278 \to 0.375 $~\\

$\sin^2 \theta_{23}  $ (NH) &~ $0.567$ ~& ~$ 0.393 \to 0.643 $~\\

$\sin^2 \theta_{23}  $ (IH) &~ $0.573$ ~& ~$ 0.403 \to 0.640 $~\\

$\sin^2 \theta_{13} $ (NH) &~ $0.0226$ ~& ~$ 0.0190 \to 0.0262 $~\\

$\sin^2 \theta_{13} $ (IH) &~ $0.0229$ ~& ~$ 0.0193 \to 0.0265 $~\\

$\delta_{\rm CP}$ (NH) & ~$1.41 \pi$ & $ ~(0 \to 2 \pi)~ $\\
$\delta_{\rm CP}$ (IH) & ~$1.48 \pi$ & $ ~(0 \to 2 \pi)~ $\\
$\Delta m_{21}^2/ 10^{-5} {\rm eV}^2 $ & $ 7.60 $ & $ 7.11 \to 8.18 $ \\

$\Delta m_{31}^2/ 10^{-3} {\rm eV}^2 ({\rm NH}) $~ &~ $ 2.48
$ & $ 2.30 \to 2.65 $ \\

$\Delta m_{31}^2/ 10^{-3} {\rm eV}^2 ({\rm IH}) $ ~&~ $ -2.38 $ & $ -2.54 \to -2.20 $ \\

\hline
\end{tabular}
\end{center}
\caption{The best-fit values and the  $3\sigma$ ranges of the neutrino oscillation parameters from Ref. \cite{ri4}. }
\label{t12}
\end{table}

\section{The Model}
The model is based on $A_4$ group \cite{r2}, which is the group of even permutation of four objects and is the smallest non-Abelian
discrete group with triplet irreducible representation. It has four irreducible representations: $1$, $1^{\prime}$, $1^{\prime \prime}$ and 3, 
with the multiplication rule
\bea
3 \times 3=1+1^{\prime}+1^{\prime \prime}+3+3\;.
\eea 
As we know, $A_4$ allows the charged-lepton mass matrix to be diagonalized by the Cabibbo-Wolfenstein matrix \cite{r7a}
\begin{equation}\label{e7}
U_{\omega}=\frac{1}{\sqrt{3}}\left( 
\begin{array}{ccc}
1&1&1\\
1&\omega&\omega^2\\
1&\omega^2&\omega
\end{array}
\right)\;,
\end{equation} 
where $\omega=e^{2 \pi i/3}=-1/2+i \sqrt{3}/2$. 

In this work, our  discussion is limited to the  leptonic sector. The  particle content of the model
includes, in addition to 
standard model fermions (i.e., the lepton doublets $l_{iL}$ and charged lepton singlets $l_{iR}$),  
three right-handed neutrinos ($\nu_{iR}$), four Higgs doublets ($\phi_i$, $\phi_0$) 
and three Higgs singlets ($\chi_i$). 
They belong to four irreducible representations of $A_4$ as given in Table \ref{t1}. 
\begin{table}[h]\label{t1}
\begin{center}
\vspace*{0.1 true in}
\begin{tabular}{|c|c|c|c|}
\hline
~~ ~&~~~$SU(2)_L$ ~~~&~~~ $U(1)_Y $~~~&~~~ $ A_4 $~~~  \tabularnewline
\hline
~~~$l_{iL}$~~~ & 2 & $-1$ &~ 3\tabularnewline
\hline

$\begin{array}{c} l_{1R}\\l_{2R}\\l_{3R} \end{array}$ & 1 & $-2$ &~$\begin{array}{c}1\\ 1^{\prime}\\ 1^{\prime \prime} \end{array}$  \tabularnewline
\hline

$\nu_{iR} $ & 1 & 0 &~ 3 \tabularnewline
\hline

$\phi_i $ & 2 & $ 1$ &~ 3\tabularnewline
\hline

$\phi_0 $ & 2  & 1 &~ 1 \tabularnewline
\hline
$\chi_i $ (real gauge singlet) & $1$ & 0 &~ 3\tabularnewline

\hline
\end{tabular}
\end{center}
\caption{Particle content of the model along with their quantum numbers.}
\label{table:t1}
\end{table}

Here $A_4$ symmetry is accompanied by an additional $U(1)_X$ symmetry as discussed in Ref. \cite{ri3}, 
which prevents the existence of Yukawa interactions of the form 
$\bar l_{iL}\nu_{iR}\tilde{\phi_i}$ and $\bar l_{iL} l_{iR}\phi_0$ 
as $l_{iL}$, $l_{iR}$,  $\tilde{\phi}_0$ have quantum number $X=1$, and all other fields have $X=0$. The phenomenologically disallowed Nambu-Goldstone 
boson does not arise in this case as $U(1)_X$ symmetry does not break spontaneously but  explicitly.
Thus,
the  Yukawa Lagrangian for  the leptonic sector is  given as \cite{r4}
\begin{eqnarray}\label{e1}
&&\mathcal{L} =-\left\{
\left[ \lambda_1 \left( \bar{l}_{iL}\phi_i\right)l_{1R}\right]+
\left[ \lambda_2\left(\bar{l}_{iL}\phi_i\right)^{\prime \prime}l_{2R}\right]+
\left[ \lambda_3\left(\bar{l}_{iL}\phi_i\right)^{\prime}l_{3R}\right] 
\right\}\\ \nonumber
&&~~~~~-\left\{\lambda_0 \left[\left( \bar{l}_{iL} \nu_{iR} \right)\tilde \phi_0\right]
+\frac{1}{2}\left[M\left( \bar{\nu}_{iR}\hat \nu_{iR}\right) \right]
+\lambda_{\chi}\left[ \left( \bar{\nu}_{iR}\hat{\nu}_{iR}\right)_3\chi_i\right]
\right\}+{\rm h.c.}\;,
\end{eqnarray}
where $\hat{\nu}_{iR}$ are  antiparticles of $\nu_{iR}$ and   $\left(\bar{l}_{iL}\phi_i\right)^{\prime}$, $\left(\bar{l}_{iL}\phi_i\right)^{\prime\prime}$ and 
$\left( \bar{\nu}_{iR}\hat{\nu}_{iR}\right)_3$ are $1^{\prime}$, $1^{\prime\prime}$ and triplet representations of $A_4$ respectively.
As the scalars $\phi_i$, $\phi_0$ and $\chi_i$ get vacuum expectation values $v_i$, $v_0$ and $\omega_i$ respectively, the above Lagrangian becomes
\begin{equation}\label{e2}
\mathcal{L}=-\bar{l}_{L}M_ll_R-\bar{\nu}_LM_D\nu_R-\frac{1}{2}\bar{\nu}_R M_R \hat \nu_R+{\rm h.c}\;,
\end{equation} 
where  $M_l$, $M_D$ and $M_R$ are charged-lepton, Dirac neutrino and right-handed neutrino mass matrices and have the forms
\begin{equation}\label{e3}
M_l=\left(\begin{array}{ccc}
\lambda_1 v_1&\lambda_2 v_1 &\lambda_3 v_1 \\
\lambda_1 v_2 &\lambda_2v_2\omega^2 &\lambda_3v_2\omega \\
\lambda_1v_3 &\lambda_2v_3\omega &\lambda_3v_3\omega^2
\end{array}
\right)\;,
\end{equation}

\begin{equation}\label{e4}
M_D=\lambda_0v_0I,\hspace*{1.5 truein}
\end{equation}
where $I$ is the identity matrix, and
\begin{equation}\label{e5}
M_R=\left( 
\begin{array}{ccc}
M&\lambda_{\chi}\omega_3&\lambda_{\chi}\omega_2\\
\lambda_{\chi}\omega_3&M&\lambda_{\chi}\omega_1\\
\lambda_{\chi}\omega_2&\lambda_{\chi}\omega_1&M
\end{array}
\right)\;.
\end{equation}
 For the vacuum alignment $v_i=v$, the charged lepton sector can be diagonalized by the transformation:
\begin{equation}\label{e6}
U_{\omega}\cdot M_l \cdot I= 
\left(
 \begin{array}{ccc}
 \sqrt{3}v\lambda_1&0&0 \\
0&\sqrt{3}v\lambda_2&0 \\
0&0&\sqrt{3}v\lambda_3
\end{array}
\right)\;,
\end{equation}
where $U_\omega$ is the Cabibbo-Wolfenstein matrix given in Eq. (\ref{e7}).
The light neutrino mass is given by the type-I seesaw formula
\begin{equation}\label{e8}
M_{\nu}=-M_D^{T}\cdot M_R^{-1} \cdot M_D\;.
\end{equation}
Since $M_D$ is proportional to an identity matrix, the neutrino mixing matrix will be the one which diagonalizes 
the right-handed neutrino mass matrix $M_R$.
The  Majorana mass matrix $M_R$ can be parameterized as
\begin{equation}\label{e9}
M_R=
\left(
\begin{array}{ccc}
A&C&D\\
C&A&B\\
D&B&A
\end{array}
\right)\;,
\end{equation}
 in a basis where charged-lepton mass matrix is not diagonal. However, in the  charged lepton mass diagonal basis 
$M_R^d=U_{\omega}^{\dagger} \cdot M_R \cdot U_{\omega}^*$ 
and can be diagonalized by tri-bimaximal (TBM)  mixing  matrix for $D=C=0$, which we don't need as it
gives vanishing $\theta_{13}$.  Even if these conditions are not satisfied 
some of the off-diagonal elements of $M_R$ become zero in TBM  basis and one can go to the TBM basis through the transformation
\begin{equation}\label{e10}
M_R^{\prime}=U_T^{\dagger}\cdot M_R \cdot U_T^*=\left(
\begin{array}{ccc}
A+B&\frac{1}{\sqrt{2}}(D+C)&0\\
\frac{1}{\sqrt{2}}(D+C)&A&\frac{i}{\sqrt{2}}(D-C)\\
0&\frac{i}{\sqrt{2}}(D-C)&B-A
\end{array}
\right)\;,
\end{equation}  
where
\begin{equation}\label{e11}
U_T=\left(
\begin{array}{ccc}
0& 1&0\\
\frac{1}{\sqrt{2}}&0&\frac{i}{\sqrt{2}}\\
\frac{1}{\sqrt{2}}&0&\frac{-i}{\sqrt{2}}
\end{array} 
\right).\hspace{1.0 truein}
\end{equation}
With the condition $D=-C$, 
$M_R^{\prime}$ becomes
\begin{equation}\label{e12}
\left(
\begin{array}{ccc}
A+B&0&0\\
0&A&i\sqrt{2}D\\
0&i\sqrt{2}D&B-A
\end{array}
\right)\;,
\end{equation}
which can be diagonalized by $U_R$, having the form
\begin{equation}\label{e13}
U_R=\left(
\begin{array}{ccc}
1&0&0\\
0&c&is\\
0&is&c
\end{array}
\right)\;,
\end{equation}
where $s$ and $c$ stand for $\sin\theta$ and $\cos\theta$ respectively and 
satisfy the relation
 \begin{equation}
\frac{cs}{c^2-s^2}=\frac{\sqrt{2}D}{B}= \frac{\sqrt{2}\omega_2}{\omega_1}\;.
\end{equation}
It should be noted that, this ratio should be real, since $\omega_{1,2}$ are VEV of real scalar fields $\chi_i$. 
The condition $C=-D$ can be realized with the vacuum alignment $\left<\chi_i\right>=(\omega_1,\omega_2,-\omega_2)$
as discussed  in \cite{r9}.
Thus,  the  lepton mixing matrix becomes
\begin{equation}\label{e14}
U=U_{\omega}\cdot U_T \cdot U_R\;,
\end{equation}
which basically known as co-bimaximal mixing matrix and predicts the mixing angles and CP violating Dirac phase as 
$\theta_{13} \neq 0$, $\theta_{23}={\pi}/{4}$ and $\delta_{CP} =\pm{\pi}/{2}$.  Also, the  mixing angles  $\theta_{12}$ and $\theta_{13}$ 
are not independent and one can express $\sin^2\theta_{12}$ in terms of $\sin^2\theta_{13}$ as
\begin{equation}
\sin^2\theta_{12}=\frac{1-3\sin^2 \theta_{13}}{3(1-\sin^2\theta_{13})},~~{\rm with}~~\sin \theta_{13} = \frac{s}{\sqrt3}\;. 
\end{equation}
To illustrate these results, we show in Fig. 1 the variation of $\sin^2 \theta_{13}$ with $\theta$ (left panel) and the correlation plot between
$\sin^2 \theta_{13}$ and $\sin^2 \theta_{12}$ (right panel). From the figure it can be seen that the observed values of 
solar ($\theta_{12}$) and reactor ($\theta_{13}$) mixing angles can be accommodated in this model.
\begin{figure}[!htb]
\includegraphics[width=7cm,height=5.0cm]{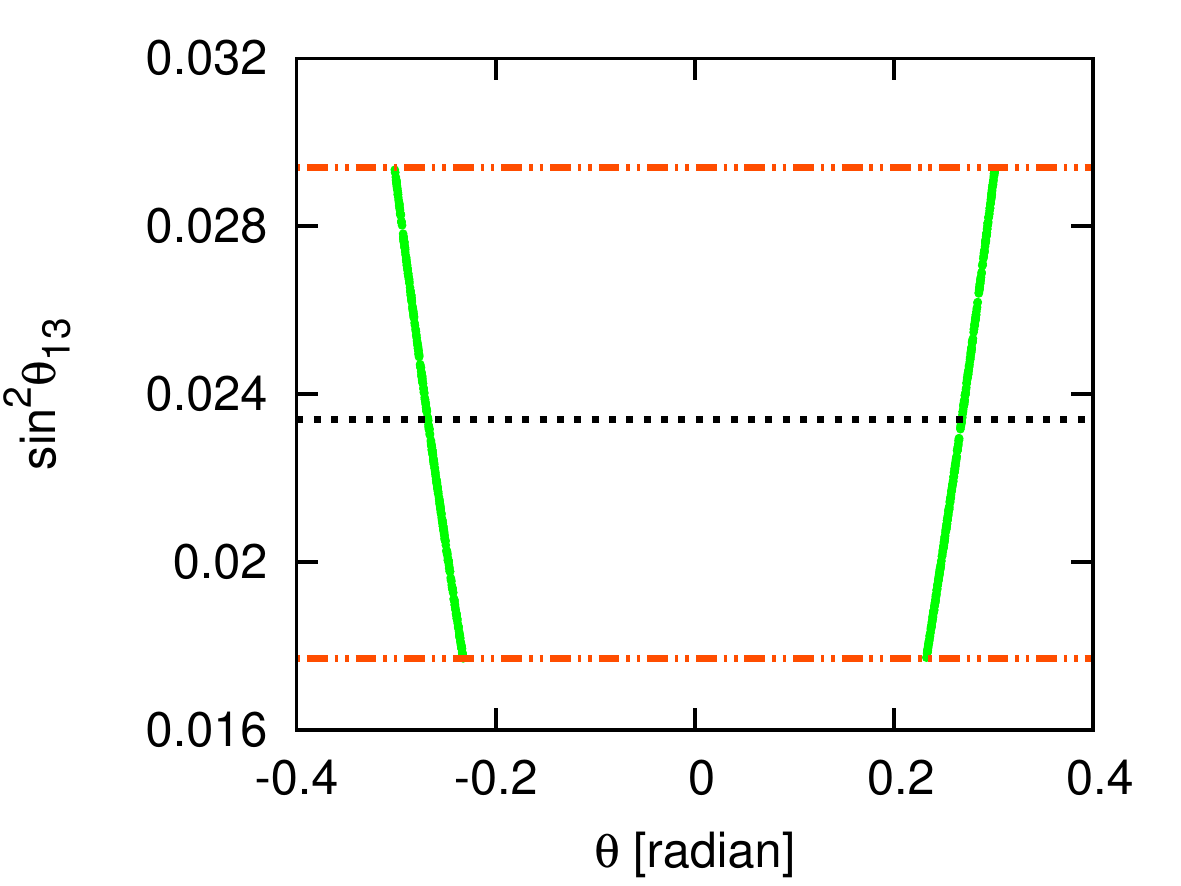}
\hspace{0.3 true in}
\includegraphics[width=7.0cm,height=5.0cm]{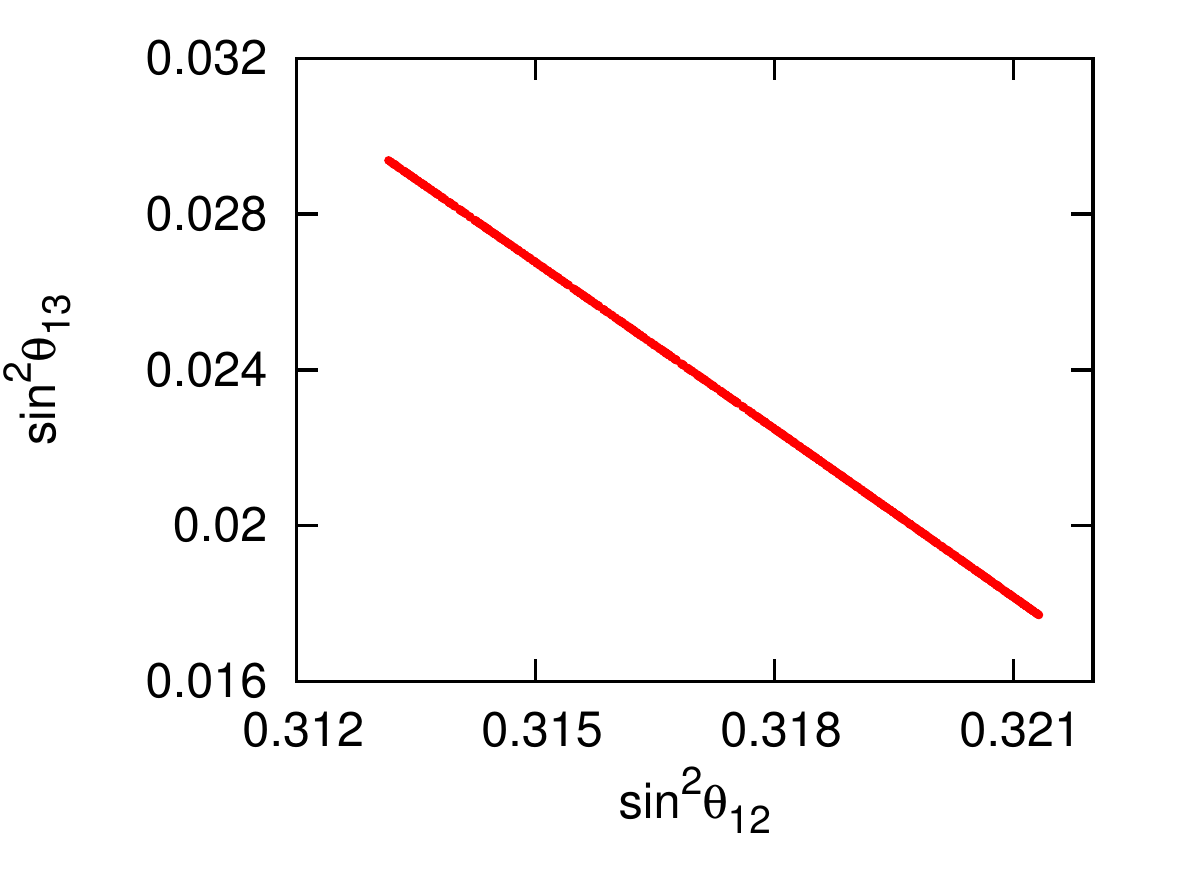}
\caption{Variation of $\sin^2 \theta_{13}$ with $\theta$ (left panel) and the correlation plots between $\sin^2 \theta_{12}$ and $\sin^2 \theta_{13}$ (right  
panel). The black dashed line in the left panel denotes the central value of $\sin^2 \theta_{13}$ and the red dot-dashed lines represent
the corresponding $3\sigma$ values.}
\end{figure}
\section{Vacuum alignment}
The complete scalar potential is given by
\begin{equation}
V=V(\phi_i)+ V(\chi_i)+V(\phi_0)+V(\phi_i \chi_i)+V(\phi_i \phi_0)
\end{equation}
with
\begin{eqnarray} 
V(\phi_i)&=&\mu_{\phi_i}^2\sum_j \phi_j^{\dagger}\phi_j +\frac{\lambda_1^{\phi_i}}{2}\left(\sum_j \phi_j^{\dagger}\phi_j\right)^2 \\ \nonumber
&+&\lambda_2^{\phi_i}\left(\phi_1^{\dagger}\phi_1 + \omega \phi_2^{\dagger}\phi_2 + \omega^2\phi_3^{\dagger}\phi_3 \right)\left(\phi_1^{\dagger}\phi_1 + \omega^2 \phi_2^{\dagger}\phi_2 + \omega \phi_3^{\dagger}\phi_3 \right) \\ \nonumber
&+& \lambda_3^{\phi_i}\left[\left( \phi_2^{\dagger}\phi_3\right)\left( \phi_3^{\dagger}\phi_2\right)+\left( \phi_3^{\dagger}\phi_1\right)\left( \phi_1^{\dagger}\phi_3\right)+\left( \phi_1^{\dagger}\phi_2\right)\left( \phi_2^{\dagger}\phi_1\right) \right]\\ \nonumber
&+& \left\lbrace \frac{\lambda_4^{\phi_i}}{2} \left[\left(\phi_2^{\dagger}\phi_3 \right)^2 + \left(\phi_3^{\dagger}\phi_1 \right)^2+\left(\phi_1^{\dagger}\phi_2 \right)^2\right]+{\rm h.c.} \right\rbrace \;,\nonumber\\
V(\chi_i)&=& \mu_{\chi_i}^2\sum_j \chi_j\chi_j +\delta^{\chi_i}\chi_1\chi_2\chi_3 +\lambda_1^{\chi_i}\left(\sum_j \chi_j \chi_j\right)^2\\ \nonumber
&+& \lambda_2^{\chi_i}\left(\chi_1\chi_1 + \omega \chi_2\chi_2 + \omega^2\chi_3\chi_3 \right)\left(\chi_1\chi_1 + \omega^2 \chi_2\chi_2 + \omega\chi_3\chi_3 \right) \\ \nonumber
&+& \lambda_3^{\chi_i}\left[\left(\chi_2\chi_3 \right)^2 + \left(\chi_3\chi_1 \right)^2 + \left(\chi_1\chi_2 \right)^2\right]\;, \nonumber\\
V(\phi_0)&=&\mu_{\phi_0}^2\phi_0^{\dagger}\phi_0 + \lambda_1^{\phi_0}\left(\phi_0^{\dagger}\phi_0\right)^2\;,
\end{eqnarray}
\begin{eqnarray}
V(\phi_i\chi_i)&=& \delta^{\phi_i\chi_i}\left( \phi_2^{\dagger}\phi_3\chi_1 + \phi_3^{\dagger}\phi_1\chi_2 +\phi_1^{\dagger}\phi_2\chi_3 \right)
+ \lambda_1^{\phi_i\chi_i}\sum_{j,k}\phi_j^{\dagger}\phi_j\chi_k\chi_k   \\ \nonumber
&+& \lambda_2^{\phi_i\chi_i}\left(\phi_1^{\dagger}\phi_1 + \omega \phi_2^{\dagger}\phi_2 + \omega^2 \phi_3^{\dagger}\phi_3\right)
\left(\chi_1\chi_1 + \omega^2 \chi_2\chi_2 + \omega \chi_3\chi_3\right) \\ \nonumber
 &+& \lambda_3^{\phi_i\chi_i}\left(\phi_2^{\dagger}\phi_3\chi_2\chi_3 +\phi_3^{\dagger}\phi_1\chi_3\chi_1+\phi_1^{\dagger}\phi_2\chi_1\chi_2 \right)+{\rm h.c.}, \nonumber
 \end{eqnarray}
\begin{eqnarray} \label{eq5}
 V(\phi_i\phi_0)&=&\lambda_1^{\phi_i\phi_0}\left(\sum_j\phi_j^{\dagger}\phi_j
 \right)\phi_0^{\dagger}\phi_0
 +\lambda_2^{\phi_i\phi_0}\left( \sum_j \phi_j^{\dagger}\phi_0~
  \phi_0^{\dagger}\phi_j\right)\\ \nonumber
 &+&\left[\lambda_3^{\phi_i\phi_0}\left(\phi_1^{\dagger}\phi_0
 \phi_2^{\dagger}\phi_3 + \phi_2^{\dagger}\phi_0\phi_3^{\dagger}\phi_1 
 +\phi_3^{\dagger}\phi_0\phi_1^{\dagger}\phi_2 \right)\right. \\ \nonumber
 &+&\left.
  \lambda_4^{\phi_i\phi_0}\left(\phi_1^{\dagger}\phi_0\phi_3^{\dagger}\phi_2 
 + \phi_2^{\dagger}\phi_0\phi_1^{\dagger}\phi_3 
 +\phi_3^{\dagger}\phi_0\phi_2^{\dagger}\phi_1  
  \right)+ {\rm h.c.} \right]\;,\\ 
V(\chi_i\phi_0)&=&\lambda^{\phi_0\chi_i}\left(\sum_j\chi_j\chi_j\right)
\phi_0^{\dagger}\phi_0.
\end{eqnarray}
The last term in Eq. (\ref{eq5}) breaks $U(1)_X$ symmetry explicitly and removes Goldstone boson which occurs due to the spontaneous breaking of $U(1)_X$ symmetry. In this model, we have the vacuum alignment $\left\langle \phi_0\right\rangle = u$, $\left\langle \phi_i\right\rangle =\left(v,v,v\right) $,  
and $\left\langle \chi_i\right\rangle =\left(w_1,w_2,-w_2\right) $ 
which is a possible minimum of scalar potential for $V(\phi_i\chi_i)=0$.   A vanishing $V(\phi_i\chi_i)$  can be achieved in the limit $\chi_i$ decouples from rest of the field as mentioned in Ref. \cite{ri3}. The decoupling of $\chi_i$ requires $\lambda_{\chi}\rightarrow 0$, $\lambda^{\phi_0\chi_i}\rightarrow0$. To generate an acceptable neutrino mass spectrum  $\lambda_{\chi}$ has to be nonzero but can be small. A small but nonzero $\lambda_{\chi}$ will generate a sufficiently small $V(\phi_i\chi_i)$ which will be too small to alter vacuum alignment considerably. In this limit the minimization condition on $u$ is given by
\begin{eqnarray}
&&\mu_{\phi_0}^2u+2\lambda_1^{\phi_0}(u^*u)u+\lambda_1^{\phi_i\phi_0}
\left(\mid v_1\mid ^2+\mid v_2 \mid ^2+\mid v_3\mid ^2\right)u+
\lambda_2^{\phi_i\phi_0}\left(\sum_{j,k}v_j^*v_k\right)u \nonumber\\
&&+{\lambda_3^{\phi_i\phi_0}}^*\left[v_1 v_2 v_3^* + v_2 v_3 v_1^*+v_3 v_1 v_2^*\right]+
{\lambda_4^{\phi_i\phi_0}}^*\left[v_1 v_3 v_2^* + v_2 v_1 v_3^*+v_3 v_2 v_1^*\right]=0\;.\label{new-26}
\end{eqnarray}
The above Eq. (\ref{new-26}) has a solution 
\begin{eqnarray}
u=\frac{{\lambda_3^{\phi_i\phi_0}}^*\left[v_1 v_2 v_3^* + v_2 v_3 v_1^*+v_3 v_1 v_2^*\right]+
{\lambda_4^{\phi_i\phi_0}}^*\left[v_1 v_3 v_2^* + v_2 v_1 v_3^*+v_3 v_2 v_1^*\right]}{\mu_{\phi_0}^2+\Big(\lambda_1^{\phi_i\phi_0}
+
\lambda_2^{\phi_i\phi_0}\Big)\left(\sum_{j}|v_j|^2\right)}
\end{eqnarray}
for $| u|^2\ll | v_i|^2$. 
\paragraph{}
Thus, for this case, i.e., for $| u|^2\ll | v_i|^2$
minimization conditions on $v_i$ are given as
\begin{eqnarray}
\frac{\partial V}{\partial v_i^*}&=&\mu_{\phi_i}^2v_i+\lambda_1^{\phi_i}v_i\sum_j| v_j|^2+\lambda_2^{\phi_i}v_i\left(2| v_i|^2-\sum_{j\neq i}| v_j|^2 \right)\\ \nonumber
&&+\lambda_3^{\phi_i}v_i\left(\sum_{j\neq i} |v_j|^2\right)+ 
\lambda_4^{\phi_i}v_i^*\sum_{j\neq i}v_j^2=0\;.
\end{eqnarray}
Considering $\lambda_4^{\phi_i}$  as real, one can get  the solution
\begin{equation}
v_i=v=\sqrt{\frac{-\mu_{\phi_i}^2}{3\lambda_1^{\phi_i}+2\left(\lambda_3^{\phi_i}+\lambda_4^{\phi_i}\right)}}\;,
\end{equation}
which is  allowed.
\paragraph{}
 Minimization conditions on $w_i$ is given by
\begin{eqnarray}
\frac{\partial V}{\partial w_1}&=& 2\left[\mu_{\chi_i}^2 + {\lambda_2^{\chi_i}}^{\prime}\left(w_2^2+w_3^2\right)\right]w_1
+\delta^{\chi_i}w_2w_3+4{\lambda_1^{\chi_i}}^{\prime}w_1^3=0\;,\\
\frac{\partial V}{\partial w_2}&=& 2\left[\mu_{\chi_i}^2 + {\lambda_2^{\chi_i}}^{\prime}\left(w_1^2+w_3^2\right)\right]w_2
+\delta^{\chi_i}w_1w_3+4{\lambda_1^{\chi_i}}^{\prime}w_2^3=0\;,\\
\frac{\partial V}{\partial w_3}&=& 2\left[\mu_{\chi_i}^2 + {\lambda_2^{\chi_i}}^{\prime}\left(w_2^2+w_1^2\right)\right]w_3
+\delta^{\chi_i}w_2w_1+4{\lambda_1^{\chi_i}}^{\prime}w_3^3=0\;,
\end{eqnarray}
one of the solutions of above set of equations is
$w_1\neq 0$, $w_3=-w_2\neq 0$, which is the vacuum alignment condition for $\langle \chi_i\rangle$.
\section{Effect of additional higgs doublets on lepton flavour violating decay $\mu \rightarrow e\gamma $}
Since $\mid u\mid^2\ll v^2$ one can neglect the mixing between $\phi_i$ and $\phi_0$ and the mass-squared matrices in the ${\rm Re}[\phi_i^0]$, ${\rm Im}[\phi_i^0]$, and $\phi_i^{\pm}$ bases have the same form \cite{rc2}.
\begin{equation}
M^2=\left(
\begin{array}{c c c}
a & b & b\\
b & a & b \\
b & b & a
\end{array}
\right),
\end{equation}
where $a=2\left(\lambda_1^{\phi_i}+2\lambda_2^{\phi_i}\right)v^2$,
  $-4\lambda_4^{\phi_i}v^2$,  $-2(\lambda_3^{\phi_i}+\lambda_4^{\phi_i})v^2$, and $b=2\left(\lambda_1^{\phi_i}-\lambda_2^{\phi_i}+
  \lambda_3^{\phi_i}+\lambda_4^{\phi_i}\right)v^2$, $2\lambda_4^{\phi_i}v^2$, $\left(\lambda_3^{\phi_i}+\lambda_4^{\phi_i}\right)v^2$ for ${\rm Re}[\phi_i^0]$, ${\rm Im}[\phi_i^0]$, and $\phi_i^{\pm}$ respectively. Hence, there are three linear combinations of $\phi_i$s, $\phi=\frac{1}{\sqrt{3}}\left(\phi_1+\phi_2+\phi_3\right)$, $\phi^{\prime}=\frac{1}{\sqrt{3}}\left(\phi_1+\omega\phi_2+\omega^2\phi_3\right)$, and $\phi^{\prime\prime}=\frac{1}{\sqrt{3}}\left(\phi_1+\omega^2\phi_2+\omega\phi_3\right)$ with vacuum expectation values $\sqrt{3}v$, $0$, and $0$ respectively. The Higgs doublet $\phi$ with mass-squared eigenvalues $\left(3\lambda_1^{\phi_i}+2\lambda_3^{\phi_i}+2\lambda_4^{\phi_i}\right)v^2$,
$0$, $0$ for ${\rm Re}[\phi^0]$, ${\rm Im}[\phi^0]$ and $\phi^{\pm}$   can be identified as standard model Higgs doublet  which gives masses to charged leptons. One can see this by expressing Yukawa interactions of $\phi_i$s with leptons in charged lepton mass diagonal basis
  \begin{eqnarray}
  \mathcal{L}&=&\left(\frac{m_e}{\sqrt{3}v}\overline{\left(\nu_e,e\right)}_Le_R+
  \frac{m_{\mu}}{\sqrt{3}v}\overline{\left(\nu_{\mu},\mu\right)}_L\mu_R+
  \frac{m_{\tau}}{\sqrt{3}v}\overline{\left(\nu_{\tau},\tau\right)}_L\tau_R\right)\phi\\ \nonumber
  &+& \left(\frac{m_e}{\sqrt{3}v}\overline{\left(\nu_{\mu},\mu\right)}_Le_R+
  \frac{m_{\mu}}{\sqrt{3}v}\overline{\left(\nu_{\tau},\tau\right)}_L\mu_R+
  \frac{m_{\tau}}{\sqrt{3}v}\overline{\left(\nu_{e},e\right)}_L\tau_R\right)\phi^{\prime} \\\nonumber
  &+&\left(\frac{m_e}{\sqrt{3}v}\overline{\left(\nu_{\tau},\tau\right)}_Le_R+
  \frac{m_{\mu}}{\sqrt{3}v}\overline{\left(\nu_{e},e\right)}_L\mu_R+
  \frac{m_{\tau}}{\sqrt{3}v}\overline{\left(\nu_{\mu},\mu\right)}_L\tau_R\right)\phi^{\prime\prime}
  \end{eqnarray}
The Higgs doublets $\phi^{\prime}$ and $\phi^{\prime\prime}$ contributes to flavour violating decays such as $\mu\rightarrow e\gamma $. The prominent contribution comes from $\phi^{\prime}$ and the branching ratio is given by
\cite{rc2},
\begin{equation}
{\rm Br}\left(\mu\rightarrow e\gamma\right)=\frac{9}{32\pi^2}m_{\tau}^4
\left(\frac{M_R^2-M_I^2}{M_R^2M_I^2}\right)^2
\left(\frac{v_0^2}{3v^2}\right)^2
\end{equation}
where $M_R^2=2\left(3\lambda_2^{\phi_i}-\lambda_3^{\phi_i}-\lambda_4^{\phi_i}\right)v^2$, $M_I^2=-6\lambda_4^{\phi_i}v^2$ are mass-squared eigenvalues of $\frac{1}{\sqrt{3}}\left({\rm Re}[\phi_1]+\omega {\rm Re}[\phi_2]+\omega^2 {\rm Re}[\phi_3]\right)$ and $\frac{1}{\sqrt{3}}\left({\rm Im}[\phi_1]+\omega {\rm Im}[\phi_2]+\omega^2 {\rm Im}[\phi_3]\right)$ respectively and  
$v_0^2=\left (1/2\sqrt{2}G_F\right )$. The predicted branching ratio will be below the experimental upper limit ${\rm Br}\left(\mu\rightarrow e\gamma\right)<4.2\times 10^{-13}$ \cite{rc1} for 
\begin{equation}
\left(\frac{M_R^2-M_I^2}{M_R^2M_I^2}\right)^{\frac{1}{2}}< 1.56\times 10^{-3}~ {\rm GeV}^{-1}\;.
\end{equation}

\section{ PERTURBATION IN NEUTRINO SECTOR}
In this section, we will consider the perturbations to mass matrices due to higher order corrections. Prominent corrections come from five-dimensional operator $\lambda_{ij}\bar{\nu}_{iR}\hat{\nu}_{jR}\chi_i\chi_j$ which modifies right-handed neutrino mass matrix. Charged lepton and Dirac neutrino masses also receive corrections from $\lambda_{jk}^{\prime}\bar{l}_{il}\phi_il_{jR}\chi_i$ and $\lambda_{jk}^{\prime}\bar{l}_{il}\tilde{\phi_0} \nu_{jR}\chi_i$ respectively, and here we are neglecting those corrections since they
allow  the mixing of  $\chi_i$ with other fields.


All elements of  Majorana mass matrix $M_R$ receive  corrections which is proportional to $\omega_1^2+\omega_2^2$ for diagonal elements and $\omega_1\omega_2$ for off diagonal elements. Since $0.04<(\omega_2/\omega_1)<0.22$, obtained from Eq. (16), using the allowed value of $s= \sqrt 3\sin \theta_{13}$, we neglect corrections to off-diagonal elements.
\begin{eqnarray}
&&\delta M_R\simeq\left(
\begin{array}{c c c}
\lambda_{11}\omega_1^2 & 0 & 0 \\
0 &\lambda_{22}\omega_1^2 &0 \\
0 &0& \lambda_{33}\omega_1^2
\end{array}
\right)\;.
\end{eqnarray}
These corrections will modify the light neutrino mass matrix and the inverse of modified light neutrino mass matrix in TBM basis can be parameterized as
\begin{equation}
M_{\nu}^{-1}=\left(
\begin{array}{c c c}
B+A&0&0\\
0 & A &i\sqrt{2}D \\
0 &i\sqrt{2}D &B-A 
\end{array}
\right)+
\left(
\begin{array}{c c c}
\frac{1}{2}\left(\lambda_{22}+\lambda_{33}\right)&0&\frac{i}{2}\left(\lambda_{33}-\lambda_{22}\right)\\
0 & \lambda_{11} &0 \\
\frac{i}{2}\left(\lambda_{33}-\lambda_{22}\right)&0&\frac{-1}{2}\left(\lambda_{33}+\lambda_{22}\right) 
\end{array}
\right)\omega_1^2\;.
\end{equation}
Hence,  in the charged lepton diagonal basis light neutrino mass matrix can be diagonalized by
\begin{equation}\label{e20}
U=U_{\omega}\cdot U_T \cdot U_R \cdot U_{13}\;,
\end{equation} 
where 
\begin{equation}
U_{13}=\left(
\begin{array}{c c c}
c^{\prime}  & 0& s^{\prime}e^{-i\phi}\\
0 & 1 & 0\\
-s^{\prime}e^{i\phi}  &0 & c^{\prime}
\end{array}
\right).
\end{equation}
with $s^{\prime}=\sin\theta^{\prime}$ and $c^{\prime}=\cos\theta^{\prime}$.

 To obtain mixing angles we compare lepton mixing matrix $U$ (\ref{e20}) with PMNS matrix (\ref{pmns}), i.e., 
\begin{equation}
U=U_{PMNS}\;.
\end{equation}
The mixing angles $\sin^2\theta_{12}$, $\sin^2\theta_{23}$ and $\sin^2\theta_{13}$ are related to the elements of $U$ as
\begin{eqnarray}\label{e21}
\sin^2\theta_{12}=\frac{|U_{12}|^2}{1-|U_{13}|^2},~ ~~\sin^2\theta_{23}=\frac{|U_{23}|^2}{1-|U_{13}|^2},~~~\sin^2\theta_{13}=|U_{13}|^2 \;,
\end{eqnarray} 
where $U_{ij}$ is the $ij^{th}$ element of the lepton mixing matrix $U$.
 Now using Eqs. (\ref{e7}), (\ref{e11}), (\ref{e20}) and (\ref{e21}), we obtain
\begin{eqnarray}
&&\sin^2\theta_{13}=\frac{1}{3}\left[ 2s^{\prime^2}-
2\sqrt{2} sc^{\prime}s^{\prime}\sin\phi +s^{2}c^{\prime 2}\right]\;,\\\label{e22a}
&&\sin^2\theta_{12}=\frac{1-s^{2}}{3-\left ( 2s^{\prime 2}-
2\sqrt{2} sc^{\prime}s^{\prime}\sin\phi +s^{^2}c^{\prime 2}\right)}\;,\\ \label{e23a}
&&\sin^2\theta_{23}=\frac{1}{2}+\frac{\sqrt{3}cc^{\prime}s^{\prime}\cos\phi}
{3-\left( 2s^{\prime^2}-
2\sqrt{2} sc^{\prime}s^{\prime}\sin\phi +s^{2}c^{\prime 2}\right)}\;,\label{e24a}
\end{eqnarray}
 Another important parameter is $J_{CP}$, the Jarlskog invariant, which is a measure of CP violation, is found to 
have the value in this model as 
\begin{eqnarray}\label{e25}
J_{CP}&=& {\rm Im}\left[U_{11}U_{22}U_{21}^*U_{12}^*\right] \nonumber\\
&=&\frac{c}{6\sqrt{3}}\left[\sqrt{2}s{c^{\prime}}^2- \left(1+c^2\right)c^{\prime}s^{\prime}\sin\phi-\sqrt{2}s{s^{\prime}}^2
\right].
\end{eqnarray}
In standard parametrization, the value of $J_{CP}$ is
\begin{equation}\label{e26}
J_{CP}=\frac{1}{8}\sin 2\theta_{12}\sin 2\theta_{23}\sin 2\theta_{13}\cos\theta_{13}\sin\delta_{CP}\;.
\end{equation}
Comparing Eqs. (\ref{e25}) and (\ref{e26}), we obtain
\begin{eqnarray}
&&\sin\delta_{CP}=\frac{\sqrt{2} s(c^{\prime 2}-s^{\prime 2})-c^{\prime}s^{\prime} (1+c^2)\sin\phi }
{\sqrt{X^\prime(2-X^\prime+s^2)\left (1-\frac{Y^{\prime 2}}{(3-X^\prime)^2} \right )}}\;,\label{27a}
\end{eqnarray}
where
\begin{eqnarray}
&&X^\prime=\left[ 2s^{\prime 2}-
2\sqrt{2} sc^{\prime}s^{\prime}\sin\phi +s^{2}c^{\prime 2} \right]\;,\nonumber\\
&&Y^\prime=2\sqrt{3}cc^{\prime}s^{\prime}\cos\phi \;.
\end{eqnarray}
\begin{figure}[!htb]
\includegraphics[width=7.0cm,height=5.0cm]{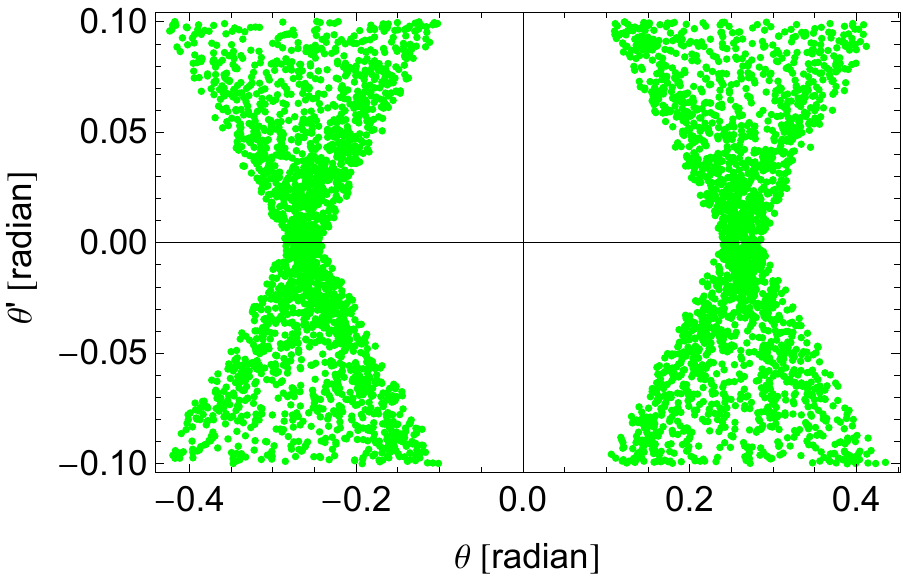}
\hspace*{0.5 truein}
\includegraphics[width=7.0cm,height=5.0cm]{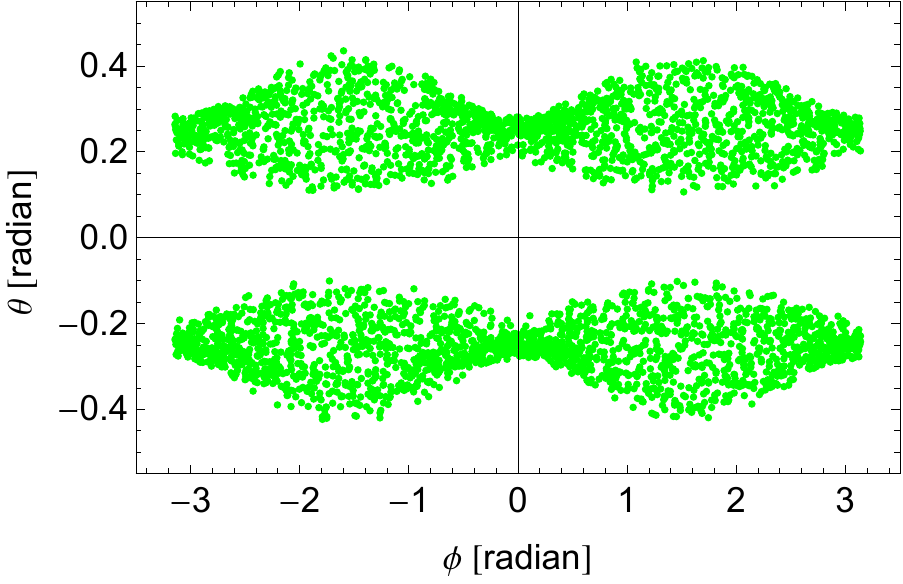}\\
\includegraphics[width=7.0cm,height=5.0cm]{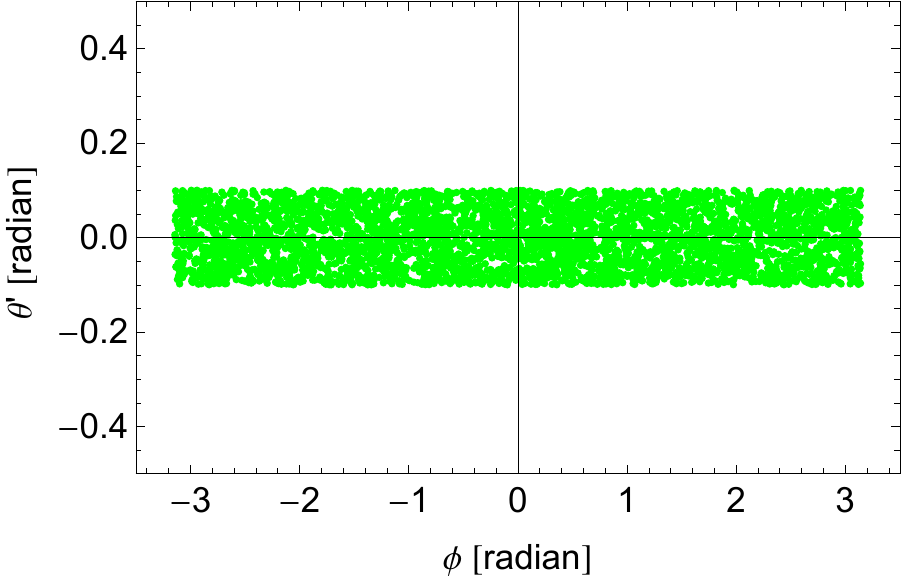}\\
\caption{Allowed parameter space in $\theta^\prime-\theta$  (left panel), $\theta-\phi$  (right panel) and $\theta^\prime - \phi$ planes compatible with the observed data.}\label{param}
\end{figure}
To show that the model predicts the mixing angles 
compatible with the observed data,we   obtain the allowed parameter space compatible with the $3 \sigma$ range of the observed data
by varying the parameters  $s$ between $[-1,1]$, $s^{\prime}$ between $[-0.1,0.1]$
   and $\phi$ between $[-\pi,\pi]$, we show the allowed parameter space in various planes  in Fig. \ref{param}. Using these allowed values of different parameters, we show the correlation plots 
between  $\sin^2 \theta_{13}$ and
$\sin^2 \theta_{23}$  (left  panel), 
 $\sin^2 \theta_{13}$ and $\sin^2 \theta_{12}$ (right  panel) and between   $\sin^2 \theta_{13}$ and $\delta_{CP}/J_{CP}$ (bottom panel)  in Fig. \ref{cor-plot}. From these plots it can be seen that by including  higher order correction to right handed neutrino mass matrix, it is possible to accommodate the observed data.  
\begin{figure}[!htb]
\includegraphics[width=7.0cm,height=5.0cm]{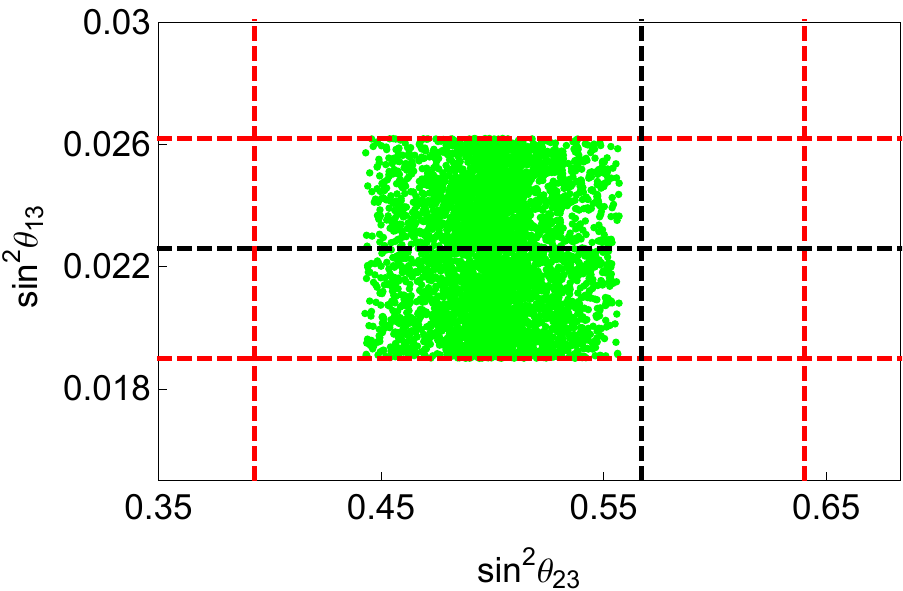}
\includegraphics[width=7.0cm,height=5.0cm]{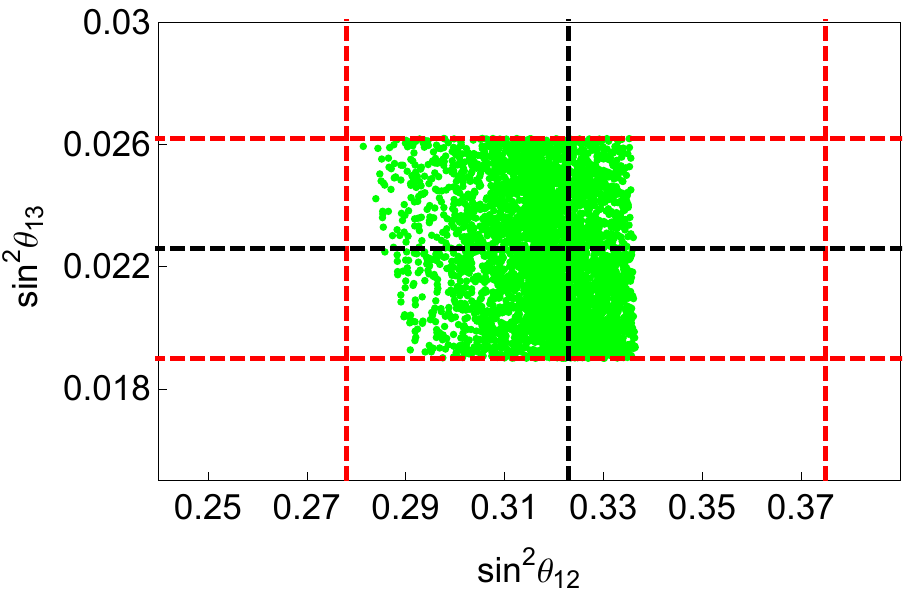}\\
\includegraphics[width=7.0cm,height=5.0cm]{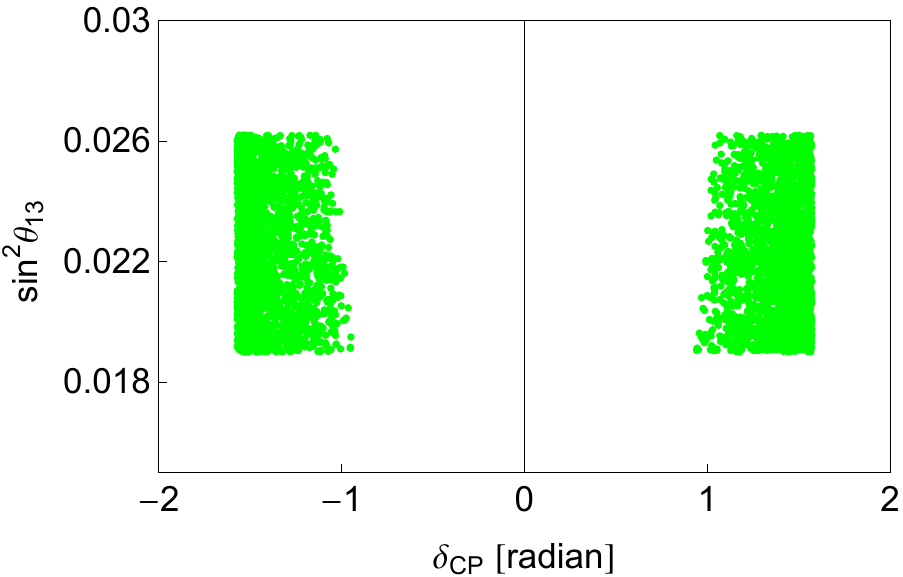}
\includegraphics[width=7.0cm,height=5.0cm]{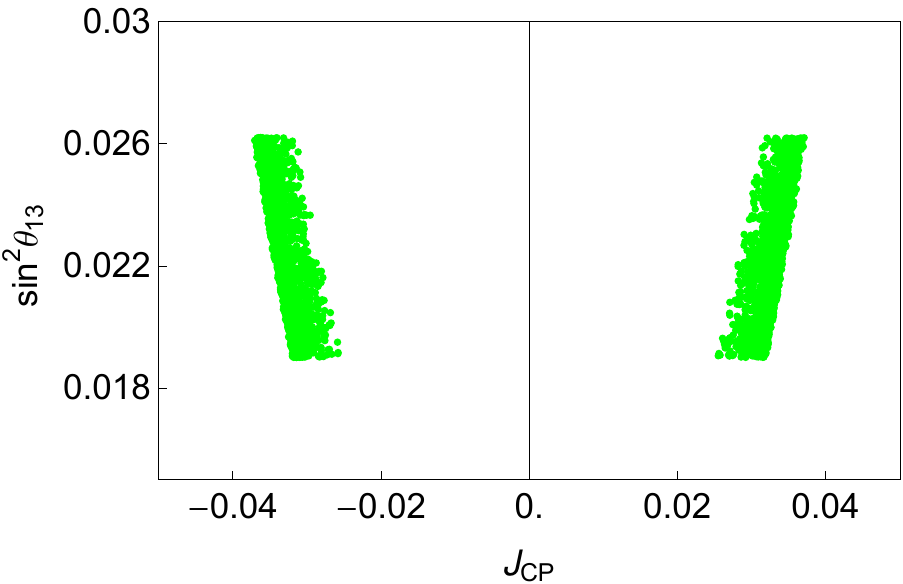}
\hspace*{0.5 truein}
\caption{Correlation plots between $\sin^2 \theta_{13}$ and
$\sin^2 \theta_{23}$  (left  panel), 
and $\sin^2 \theta_{13}$ and $\sin^2 \theta_{12}$ (right  panel) and between   $\sin^2 \theta_{13}$ and $\delta_{CP}/J_{CP}$ (bottom panel) including the corrections.}\label{cor-plot}
\end{figure}

 In this model, light neutrinos acquire Majorana masses through Type-I seesaw which indicates neutrinos are of Majorana type. 
Majorana nature of neutrinos predict the  existence of neutrino-less double beta decay ($0\nu \beta\beta$), 
which is a process where two neutrons inside a nucleus 
convert into two protons without emitting neutrinos, i.e., $(A,Z)\rightarrow(A,Z+2)+2e $.
Several experiments like KamLAND-Ze \cite{r11}, EXO \cite{r12} and GERDA \cite{r13} are searching for the  neutrino-less double beta decay. 
These experiments put upper bound on $|M_{ee}|$, the (1,1) element of neutrino mass matrix,
  since the half-life of $0\nu \beta\beta$ decay is proportional 
to $|M_{ee}|^2$. The expression for $|M_{ee}|$  in the flavor basis is
\begin{equation}
|M_{ee}|=| U_{11}^2 m_1+U_{12}^2 m_2+ U_{13}^2 m_3|\;,\label{Mee}
\end{equation}
where $m_1$, $m_2$, and $m_3$ are light neutrino masses and $U_{1j}$'s are elements of first row of the lepton mixing matrix $ U$,
which are given as
\begin{eqnarray}
U_{11}&=& \frac{2}{\sqrt{6}}c^{\prime}-\frac{i}{\sqrt{3}}ss^{\prime}e^{i\phi}\;, \nonumber\\
U_{12} &=& \frac{1}{\sqrt{3}}c\;, \nonumber\\
U_{13}&=& \frac{2}{\sqrt{6}}s^{\prime}e^{-i\phi}+\frac{i}{\sqrt{3}}sc^{\prime}\; . 
\end{eqnarray}
The lowest upper bound on $|M_{ee}|$ is 0.22 eV came from GERDA phase-I data. Here we study the variation of $|M_{ee}|$ 
with the lightest neutrino mass $m_1~(m_3)$, in the case of normal (inverted) hierarchy as shown in Fig. 4. In our calculation we have used 
the relations
\begin{eqnarray}
m_2&=&\sqrt{m_1^2+\Delta m_{21}^2}\;,\nonumber \\
m_3&=&\sqrt{m_1^2+\Delta m_{31}^2}\;,
\end{eqnarray}
for normal hierarchy and
 \begin{eqnarray} 
 m_1 &=&\sqrt{m_3^2+\Delta m_{13}^2}\;, \nonumber\\
 m_2 &=&\sqrt{m_3^2+\Delta m_{13}^2+\Delta m_{21}^2}\;,
 \end{eqnarray}
 for inverted hierarchy,
 and obtained upper limit on $m_1$ ($m_3$) as 0.071 (0.065 eV)  taking into account the  cosmological upper bound on $\Sigma_im_i$ as 0.23 eV \cite{r10}.
Another observable is   the kinetic electron neutrino mass
in beta decay ($m_e$), which is probed in direct search
for neutrino masses, can be expressed as
\begin{equation}
 m_{e}=\sqrt{ |U_{11}|^2 m_1^2 +|U_{12}|^2 m_2^2+ |U_{13}|^2 m_3^2}\;.
 \label{me}
\end{equation}

In the right panel of Fig. 4, we show the variation of $m_e$ with the lightest neutrino mass $m_1$ ($m_3$) for normal hierarchy (inverted hierarchy) case,
and the upper limit on $m_e$ is found to be 0.07 (0.08) eV.
\begin{figure}[!htb]
\includegraphics[width=7.0cm,height=5.0cm]{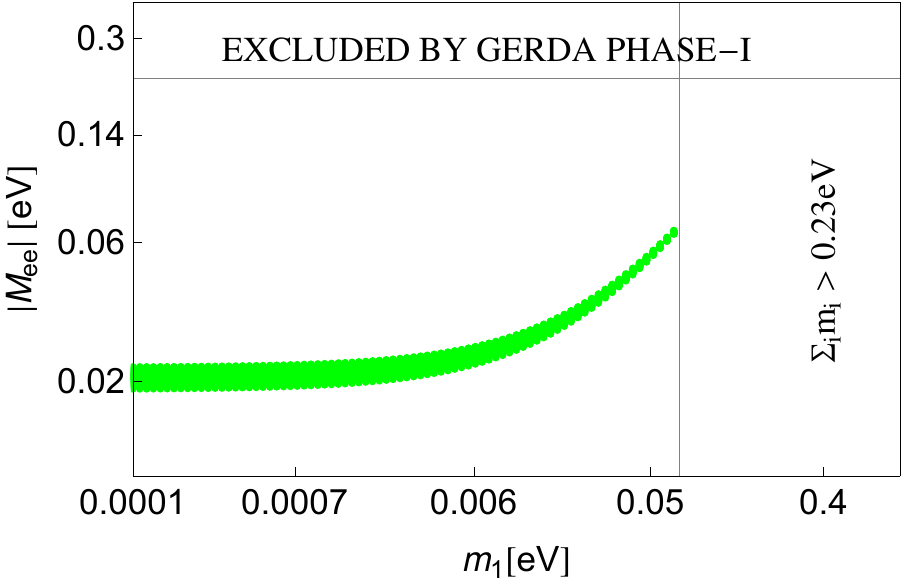}
\hspace*{0.5 truein}
\includegraphics[width=7.0cm,height=5.0cm]{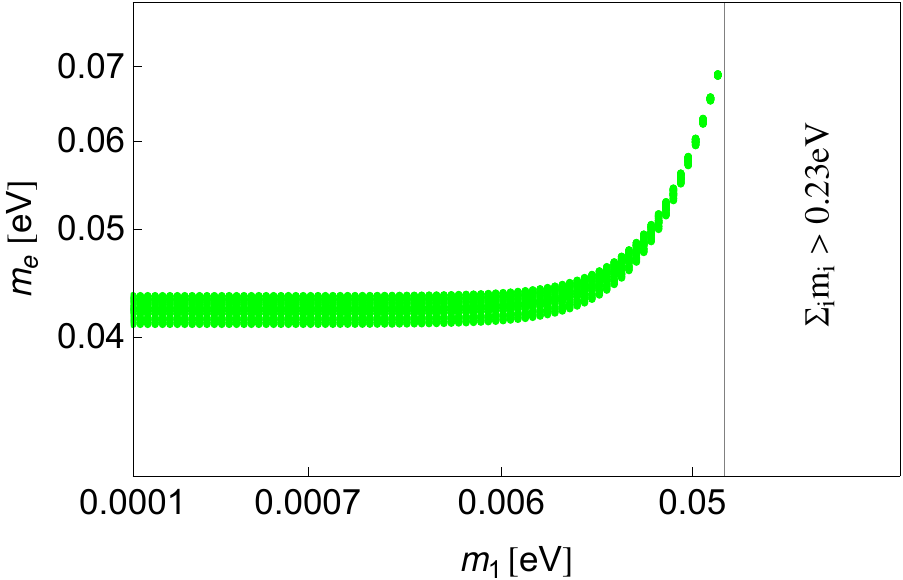}\\
\includegraphics[width=7.0cm,height=5.0cm]{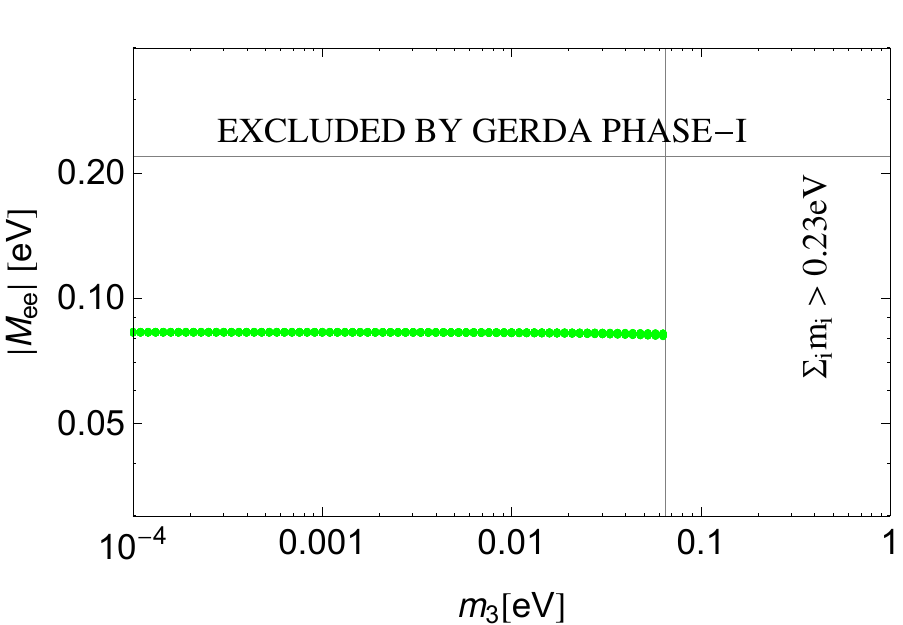}
\hspace*{0.5 truein}
\includegraphics[width=7.0cm,height=5.0cm]{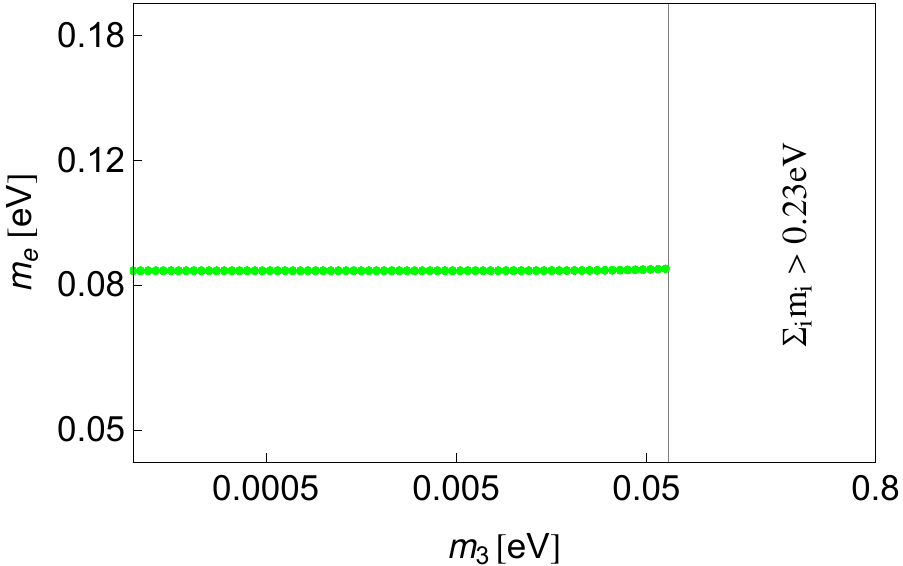}\\
\caption{Variation of $|M_{ee}|$ with the lightest neutrino mass $m_1$ $(m_3)$ (left panel) and $m_e$ vs. $m_1$ $(m_3)$ in the right panel
 for the normal mass hierarchy (inverted mass hierarchy) case.}
\label{6a}
\end{figure}

\section{Conclusions}
We consider a model based on $A_4$ symmetry, which gives co-bimaximal form ($\theta_{23}={\pi}/{4}$, $\delta_{CP}=\pm{\pi}/{2}$ 
and $\theta_{13}\neq0$) for the leading order neutrino mixing matrix. 
 There are four Higgs doublets $\phi_0$, and $\phi_i$, for $i=1,2,3$ in this model. One of the three linear combinations  ($\phi$) of $\phi_i$  behaves exactly as standard model Higgs doublet while neutral component of the other two ($\phi^{\prime}$, $\phi^{\prime\prime}$) contribute to  the lepton flavour violating decays such as $\mu\rightarrow e \gamma$. 
We have considered higher order corrections in neutrino sector 
coming from five-dimensional  operators after spontaneous breaking of $A_4$ symmetry.
The mixing angles, thus obtained are found to be within 
the $3\sigma$  ranges of their experimental values. The CP violating phase $\delta_{CP}$ is found to be around the region $\pm\pi/2$,
and the upper limit on the Jarlskog invariant is ${\cal O}(10^{-2})$. We also studied the variation of the effective neutrino mass $|M_{ee}|$  with the 
lightest neutrino mass $m_1$ ($m_3$) in the case of normal (inverted) hierarchy and found  its value  to be lower than the experimental upper 
limit for all allowed values of $m_1$ ($m_3$).


{\bf Acknowledgments}
SM  would like to thank University Grants Commission for financial support.
The work of RM was partly supported by the  Science and Engineering Research Board (SERB),
Government of India through grant No. SB/S2/HEP-017/2013.

\end{document}